\documentclass[
    ,final            
  ]
  {aipproc}

\layoutstyle{8x11double}

\usepackage{amssymb}

\newcommand{\be}{\begin{eqnarray}}
\newcommand{\ee}{\end{eqnarray}}

\newcommand{\ma}{\Delta m^2_{31}}
\newcommand{\chr}{\mbox{$\breve{\rm C}$erenkov~}}

\newcommand{\stch}{\sin^2 2\theta_{13}}

\newcommand{\sgnma}{\mathrm{sgn}(\Delta m^2_{31})}

\newcommand{\bb}{$\beta$-Beam~}

\newcommand{\br}{$^8$B~}
\newcommand{\li}{$^8$Li~}
\newcommand{\he}{$^6$He~}
\newcommand{\neon}{$^{18}$Ne~}

\begin{document}

\title{Optimization of the Two-Baseline Beta-Beam}

\classification{13.15.+g, 13.90.+i,   14.60.Lm,  14.60.Pq}
\keywords      {Neutrino oscillations, $\beta$-Beam}

\author{Pilar Coloma\thanks{email: p.coloma@uam.es}}{ 
  address={\begin{center}Instituto de F\'isica Te\'orica, Universidad Aut\'onoma de Madrid,\\ 28049 Madrid, Espa\~na \end{center}}
}

\begin{abstract}
We propose a $\beta$-Beam experiment with four source ions and two baselines for the best possible sensitivity to $\theta_{13}$, CP violation and mass hierarchy. Neutrinos from \neon and \he with Lorentz boost $\gamma=350$ are detected in a 500 kton water \chr detector at a distance $L=650$ km from the source. Neutrinos from \br and \li are detected in a 50 kton magnetized iron detector at a distance $L=7000$ km from the source. Since a tilt angle $\vartheta=34.5^\circ$ is required to send the beam to the magic baseline, the far end of the ring has a maximum depth of $d=2132$ m. We alleviate this problem by proposing to trade reduction of the decay ring with the increase in the boost factor of the \li and \br ions up to $\gamma_{^8\mathrm{Li}}=390$ and $\gamma_{^8\mathrm{Li}}=650$, such that the number of events at the detector remains almost the same. We study the sensitivity reach of this two-baseline two-storage ring $\beta$-Beam experiment, and compare it with the corresponding reach of the other proposed facilities.

\end{abstract}

\maketitle


\section{Introduction}

If the next-generation reactor- and accelerator-based experiments fail to observe any positive signal for non-zero $\theta_{13}$, more powerful experiments will be needed. In addition to this, even if the forthcoming experiments could measure $\theta_{13}$, it is quite unlikely that they will obtain neither a CP violating signal nor a measurement of $\sgnma$. Therefore, a new generation of experiments will be needed also in this case. Two main options have been envisaged for this purpose: the Neutrino Factory \cite{NF}, and the so-called ``$\beta$-Beam'' \cite{zucc}. While the former produces the (anti)neutrino beams through muon decay, the latter entails producing $\beta$-unstable radioactive ions and letting them decay in a storage ring.

We propose a \bb set-up  where we produce, accelerate and store ions of the four kinds (\he, \neon, \li and \br$\!\!\!\!$) at CERN, each of them running for a period of 2.5 years. We aim the \he- and \neon-generated low-energy neutrino beams to a megaton water \chr detector located at $L=650$ km from the source, and the \li- and \br-generated high-energy neutrino beams to a 50 kton iron detector at a distance close to the magic baseline. We have also reconsidered the storage ring design and feasibility: as we consider different ions for the two baselines, we have reduced the size of the ring that is aiming at the far detector, increasing at the same time the $\gamma$ factor for the \br and \li ions in order not to lose sensitivity to the main observables we are interested in.

\section{Two-Baseline $\beta$-Beam Experiment}

\subsection{The choice of the two baselines}

It is well-known that for a baseline of \emph{O}~$\sim 7000 \textrm{km} $, all the $\delta$-dependent terms in the golden channel probability vanish~\cite{magic2,BurguetCastell:2001ez}. This provides a clean bedrock to determine $\theta_{13}$, because the intrinsic degeneracy disappears. In addition to this, the density encountered by the (anti)neutrinos at this baseline allows for a resonant enhancement in the probability when $E_\nu \sim 6$~GeV if the mass hierarchy is normal (inverted). The advantage of this resonance is two-fold: first, it compensates the loss of events due to the very long baseline; second, it only occurs for (anti)neutrinos if the mass hierarchy is normal (inverted), therefore providing an extremely good probe of the mass ordering. In order to reach this range of energies, \br and \li ions are the best candidates because of their high end point energies.

However, $\delta$ cannot be measured at the magic baseline, and matter effects can fake true CP violation. Therefore, short baselines are better if we want to probe $\delta$. In the small matter effect regime, maximizing the CP-violating terms translates into fixing $L/E=515$ km/GeV. The mean neutrino energy of neutrinos coming from $^6$He and $^{18}$Ne decays at $\gamma = 350$ is $E_0 \gamma \sim 1.2$~GeV, which translates to an on-peak baseline of $L=618$~km, while for \br and \li, due to their much higher end point energies, this baseline would be $L\sim 1500-2000$ km. The main problem in this case is that the flux is proportional to $L^{-2}$. Therefore, \he and \neon turn out to be the better candidates to probe CP violation.

\subsection{The Storage Ring}

The original design of the storage ring proposed in \cite{zucc} must be modified when the boost factor $\gamma$ is increased. If we use LHC dipolar magnets (with a maximum magnetic field of 8.3 T) to bend the ions, and keeping the straight sections untouched, the useful fraction of ion decays (also called ``livetime") for this ring would be\footnote{$l_{\rm racetrack} = \frac{L_{\rm straight}}{2 L_{\rm straight} + 2 \pi R}$} {\em l} = 0.28, the total length of the decay ring being $L_r = 8974$ m. The tilt angle needed to aim at a detector located at 650 km from the source is $\vartheta = 3^\circ$: this means that the maximum depth of the far end of the ring is $d = 197$ m.

However, if a ring of the same type is used to aim at a detector located at $L= 7000$ km from the source, the tilt angle to be considered is $\vartheta = 34.5^\circ$. In this case, the maximum depth of the far end of the  ring is $d = 2132$ m, something well beyond any realistic possibility. In our proposal, two different storage rings will be used to aim to the detectors\footnote{There is no particular advantage except budget in using the same ring to aim at both sites simultaneously.}. Therefore, it is possible to design two rings of different characteristics and reduce the size of the ring that is aiming at the far detector.

Note that with the refurbished SPS (SPS+) the \br and \li ions could be accelerated up to $\gamma = 650$ and $\gamma=390$, respectively. Due to the resonance, a $10\%$ increase in the $\gamma$ factor for \li produces an increase of a $40\%$ in the number of antineutrino events at the detector. Therefore, we can use a ring with a much shorter straight section, $L_s = 998$ m and the physics reach of the setup will remain practically unaffected, in spite of the fact that it will have a lower livetime {\em l} = $0.6 \times 0.28 \sim 0.17$. The maximum depth this ring would reach is $d = 1282$ m, which is still much larger than what is needed for the Neutrino Factory but is almost 1 km less deep than in the original design.

\subsection{The Detectors}

We make the following choices for our detectors: (1) Since CP measurements are better at lower energies with \neon and \he as source ions \cite{twobaseline1,twobaseline2}, it is preferable to have a detector with lower threshold and good energy resolution. We opt for a water \chr detector with 500 kton fiducial mass (as in Refs.~\cite{bc,bc2}). This detector could be housed at Canfranc, for example, at a distance of 650 km from the $\beta$-Beam at CERN; (2) Mass hierarchy measurement is the main motivation for the experiment at the magic baseline, for which higher energy neutrinos from highly boosted \br and \li ions will be used. We prefer thus to use a 50 kton magnetized iron detector at this baseline. This far detector could be the ICAL@INO detector in India \cite{ino} which is at a distance of 7152 km, and which will soon go under construction.

The efficiencies and beam-induced backgrounds expected in a water \chr detector for the $\gamma = 350$ $\beta$-Beam fluxes from \neon and \he decays are given in \cite{bc2} as migration matrices that we use to simulate the detector. 
Unfortunately, a similarly detailed analysis of the performance of the iron detector exposed to the $\beta$-Beam fluxes is lacking. We therefore follow the efficiencies and backgrounds derived in \cite{Abe:2007bi} for the Neutrino Factory fluxes instead. Notice that this is a very conservative assumption, since for a \bb charge identification is not mandatory, unlike in the NF. Therefore, we could take advantage of this and use the charge ID capability of the detector to further reduce the background. 

Finally, note that the largest uncertainties in the performance of the iron detector are on the efficiencies and backgrounds for the events of lowest energy, around $1-5$ GeV. However, the main role of the iron detector is to observe the resonance in the probability, which takes place around $6-7$ GeV, so these uncertainties will practically have no effect at all in the performance of the setup.

\section{Comparative Sensitivity Reach}

We quantify the sensitivity reach of the experiments in terms of three different performance indicators: 
\begin{enumerate}
\item The $\stch$ discovery reach: This is the minimum true value of $\stch$ for which the experiment can rule out at 
the 1 d.o.f. $3\sigma$ the value $\stch=0$ in the fit, after marginalizing over all the other parameters. 
\item The CP-violation 
reach: This is the range of $\delta$ as a function of $\stch$ which can rule out no CP-violation ($\delta=0$ and $180^\circ$) at the 1 d.o.f. $3\sigma$, after marginalizing over all the other parameters. 
\item The $sgn(\ma)$ reach in $\stch$: This is defined as the limiting value of $\stch$ for which the wrong hierarchy can be 
eliminated at $3\sigma$. 
\end{enumerate}

\subsection{Comparison with previous \bb proposals}

\begin{figure}\begin{tabular}{cc}
  \includegraphics[width=0.40\textwidth,angle=0]{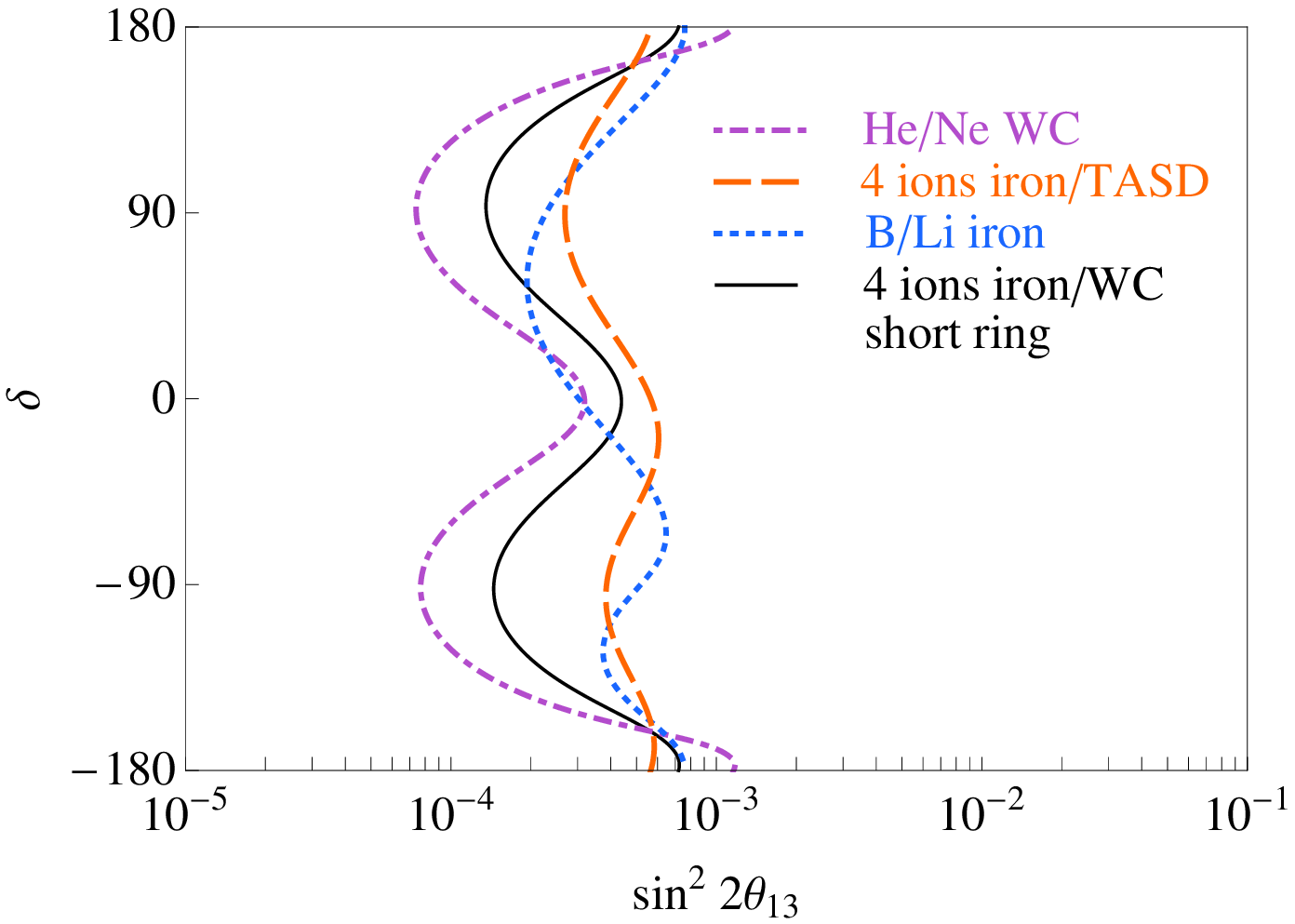} &
  \includegraphics[width=0.40\textwidth,angle=0]{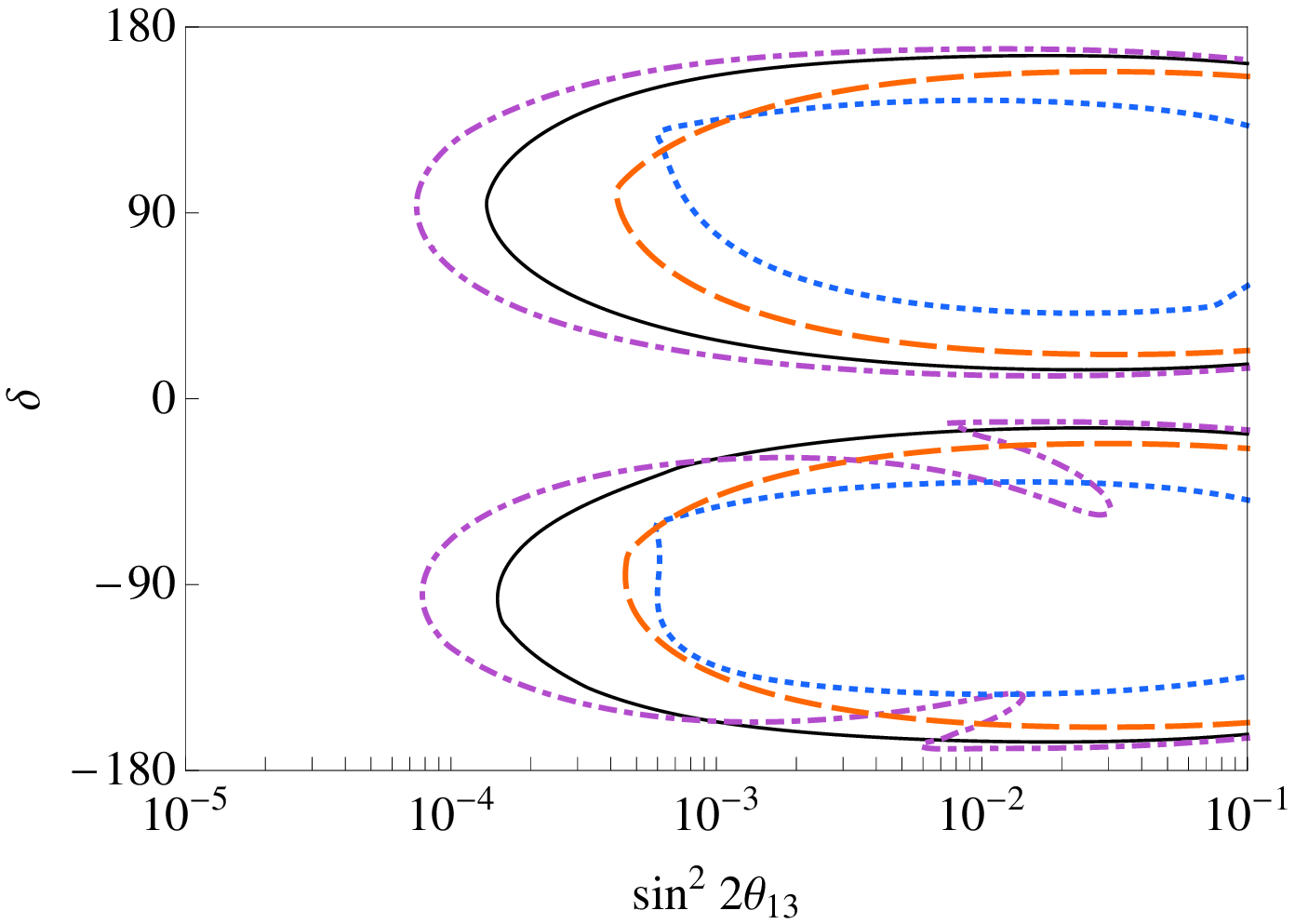} \\
  \includegraphics[width=0.40\textwidth,angle=0]{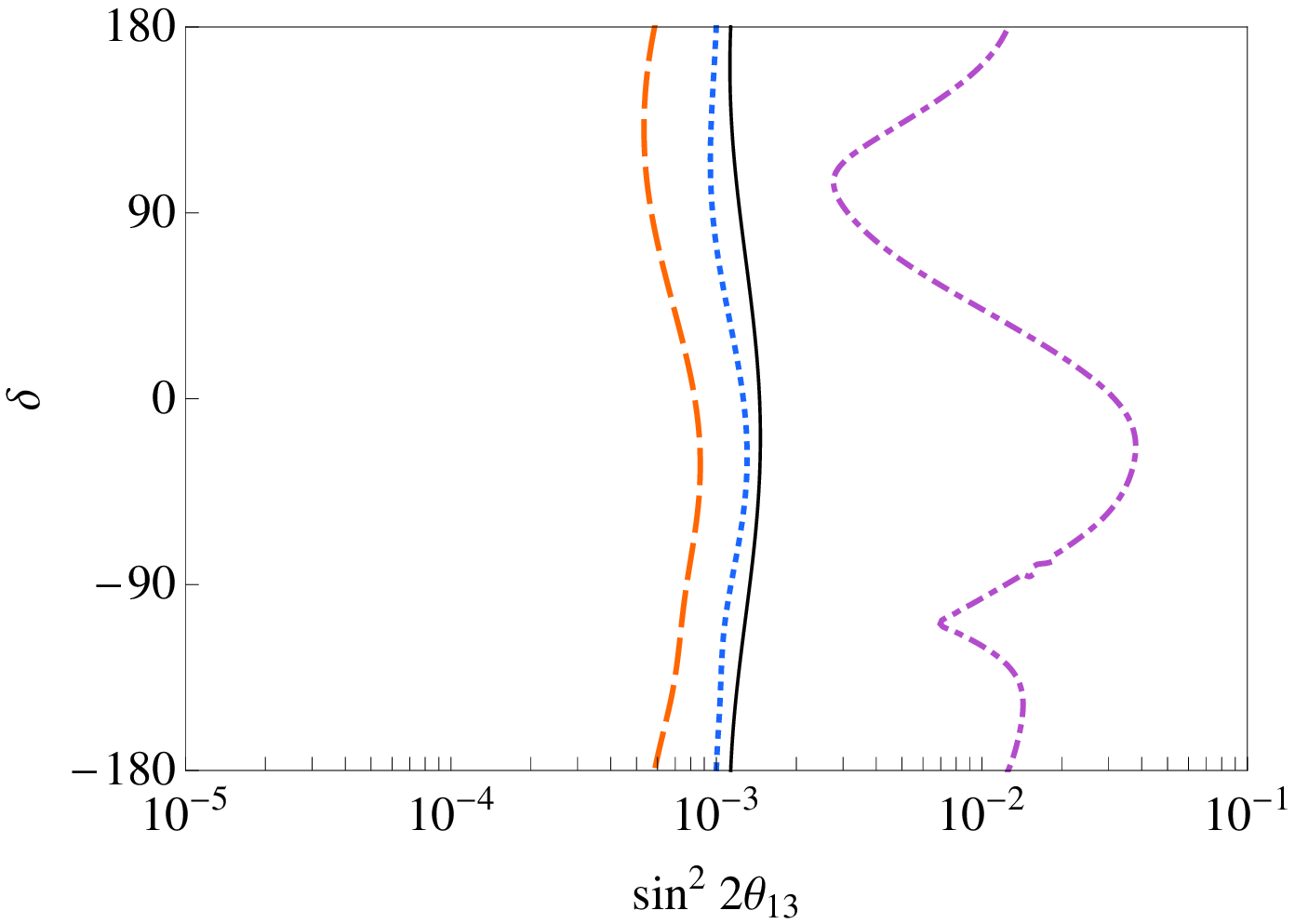} &
  \includegraphics[width=0.40\textwidth,angle=0]{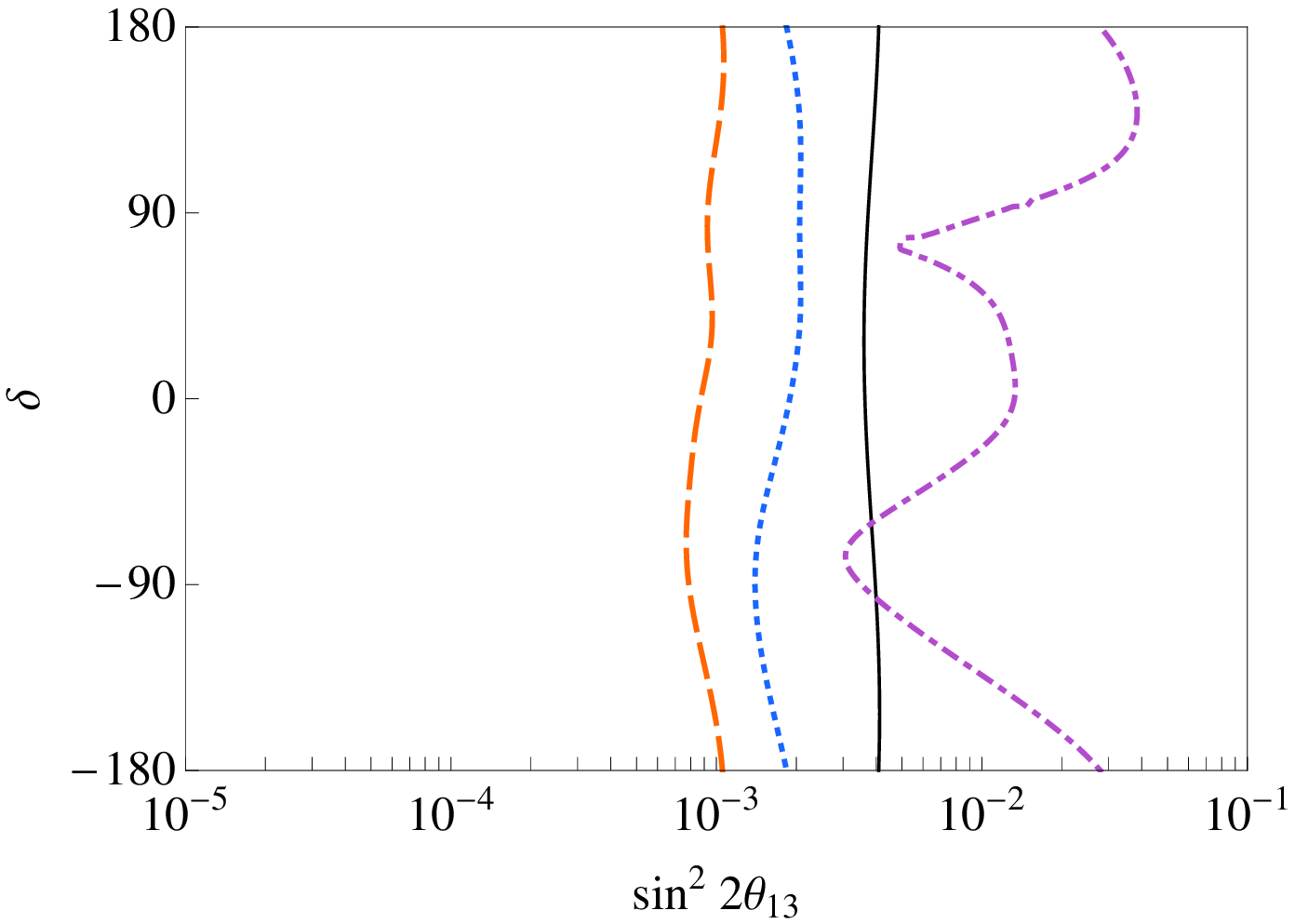} 
\end{tabular}
  \caption{\label{fig:sens} Sensitivity reach of the different \bb set-ups in terms for the three performance indicators defined in the text. The upper left hand panel shows $\stch$ discovery reach, the upper right hand panels shows the CP-violation reach, while the lower panels show the mass hierarchy discovery reach for normal (left panel) and inverted (right panel) hierarchy. The different line types are for different \bb set-ups as described in the text.}
\end{figure}

In Fig. \ref{fig:sens} we present the comparison of our set-up with respect to other high $\gamma$ \bb proposals, defined as follows: 

\begin{enumerate}
\item Solid, black lines: These correspond to our proposal~\cite{twobaseline}. 

\item Blue, dotted lines: The two-baseline \bb set-up proposed in~\cite{twobaseline1}. Here neutrino beams from decay of \br and \li with boost factor $\gamma = 350$, are detected in two 50 kton magnetized iron detector located at 2000 km and 7000 km respectively.

\item Orange, dashed lines: The two-baseline \bb set-up proposed in~\cite{twobaseline2}. Here all four ions are used. Beams 
from decays of \neon and \he accelerated to $\gamma = 575$ are detected in a 50 kton Totally Active Scintillator Detector (TASD) at 730 km. Beams from decays of \br and \li accelerated to $\gamma = 656$ are detected in a 50 kton magnetized iron detector at 7000 km. 

\item Purple, dot-dashed lines: The one-baseline \bb set-up proposed in~\cite{bc,bc2}. Neutrino beams produced by \neon and \he decays, each accelerated to $\gamma =350$ are detected in a 500 kton water \chr detector located at 650 km. 
\end{enumerate}

The original $\beta$-Beam proposals assumed "standard" useful fluxes of $1.1\times 10^{18}$ and $2.9\times 10^{18}$ decays per year for \neon and \he respectively. Similar standard numbers regarding \br and \li are lacking. At present, \li and \he can be easily produced with the standard ISOLDE techniques. \br is easily produced, but it tends to interact with the medium nearby. Finally, \neon production is still difficult, and we are far from the original goal of $1.1\times 10^{18}$ ions per year.

Preliminary studies on the production rates of \br and \li show that it could be enhanced in the near future. $\beta$-Beams  are facilities under study for construction in the next two decades. Therefore, we will assume that $10^{19}$ ions per year can be stored into the ring, for all ion species\cite{Donini:2008zz}. Note that for the ``standard'' storage ring considered in set-ups 2, 3 and 4, the livetime is {\em l} = 0.28, what translates into {\it $3 \times 10^{18}$ useful decays per year} per polarity.
However, for our proposal, the storage ring for the \br and \li ions is shorter, giving a 40\% smaller livetime. 

We have considered 2.5\% and 5\% systematic errors on the signal and on the beam-induced background, respectively. They have been included as ``pulls'' in the statistical $\chi^2$ analysis. The following $1 \sigma$ errors for the oscillation parameters were also considered: 
$\delta \theta_{12} = 1\%$, $\delta \theta_{23} = 5\%$, 
$\delta \Delta m^2_{21} = 1 \%$ and $\Delta m^2_{31} = 2\%$. 
Eventually, an error $\delta A = 5\%$ has been considered for the Earth density given by the PREM model \cite{prem}. 
Marginalization over these parameters has been performed for all observables. The Globes 3.0 \cite{globes1,globes2} software was 
used to perform the numerical analysis.

The upper left hand panel of Fig.~\ref{fig:sens} shows the $\stch$ discovery reach. It can be seen that all the set-ups perform in a similar way. However, notice that the setups which have a far detector at the magic baseline show less $\delta$ dependence, while setup 4 shows the strongest. In spite of the far detector, our proposal also shows some $\delta$ dependence, because the near detector in this case is 10 times larger than the far one.

The upper right hand panel shows the CP-violation discovery potential. This is best at the shorter baselines. Thus, set-up 4, from \cite{bc,bc2} has sensitivity to CP-violation for the smallest values of $\stch$, since the near detector is exposed to the beam for ten years. However, as this setup does not have a far detector it cannot determine the mass hierarchy, and a loss of sensitivity appears for negative values of $\delta$ around $\sin^2 2\theta_{13}\sim 10^{-2}$, due to the sign degeneracy. In our setup, however, the far detector provides sensitivity to the mass hierarchy for values of $\sin^2 2\theta_{13}$ down to $10^{-3}$ ($3\times10^{-3}$) for normal (inverted) hierarchy independently of the value of $\delta$, therefore solving these degeneracies.

The lower panels show the sensitivity to the mass hierarchy. This is best at the far detectors and thus, the best sensitivities are achieved for set-up 3 from \cite{twobaseline2} due to the higher statistics granted by the larger gamma factor assumed of $\gamma = 656$ for both \br and \li. Since for the set-up we propose here we restrict to the maximum $\gamma$ attainable at the SPS+, which for \li is $\gamma = 390$, the difference between the two set-ups is larger for the inverted hierarchy (lower right hand panel).

\subsection{Comparison with the Neutrino Factory}

\begin{figure}[t]\begin{tabular}{cc}
\includegraphics[width=0.40\textwidth,angle=0]{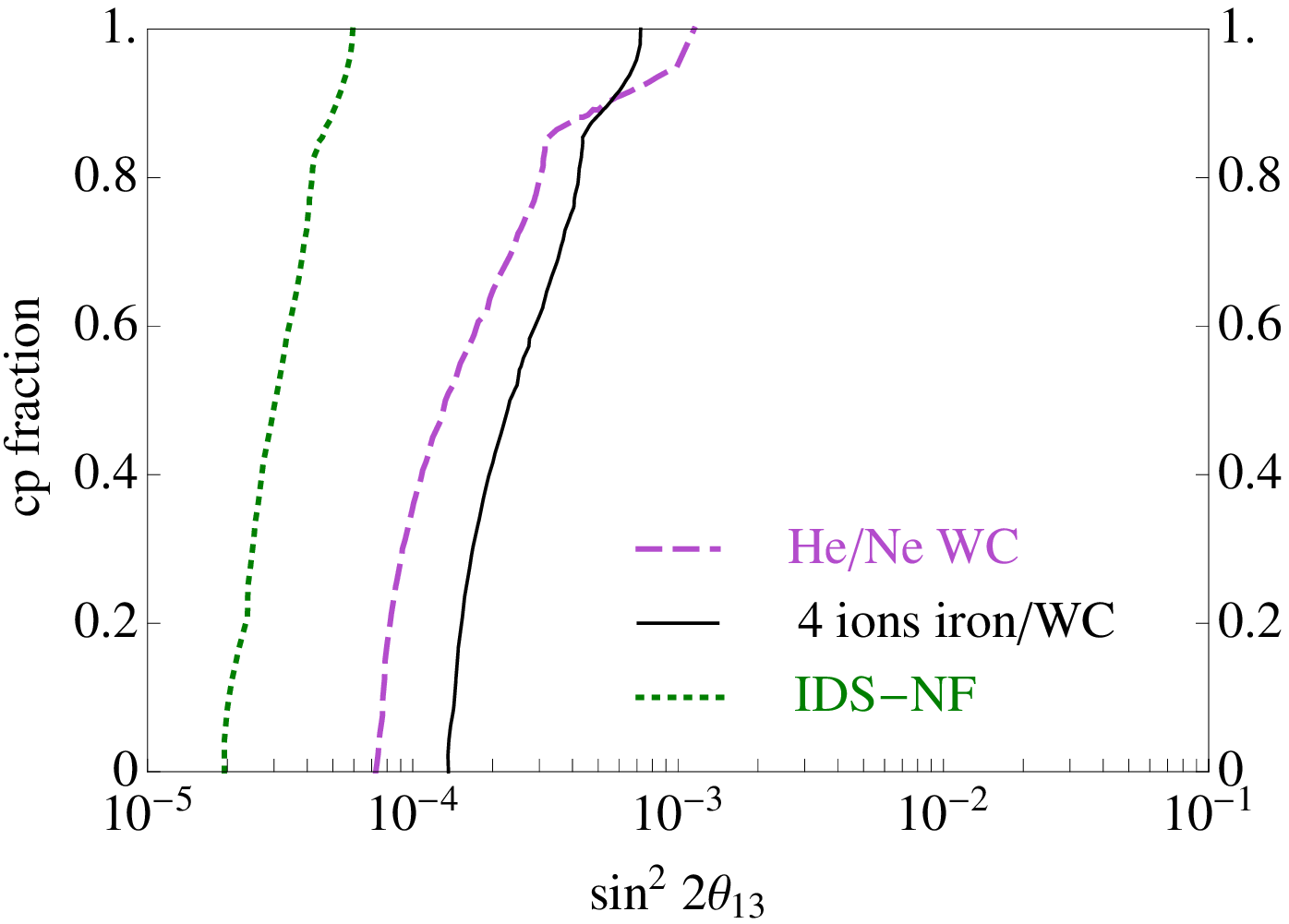}&
\includegraphics[width=0.40\textwidth,angle=0]{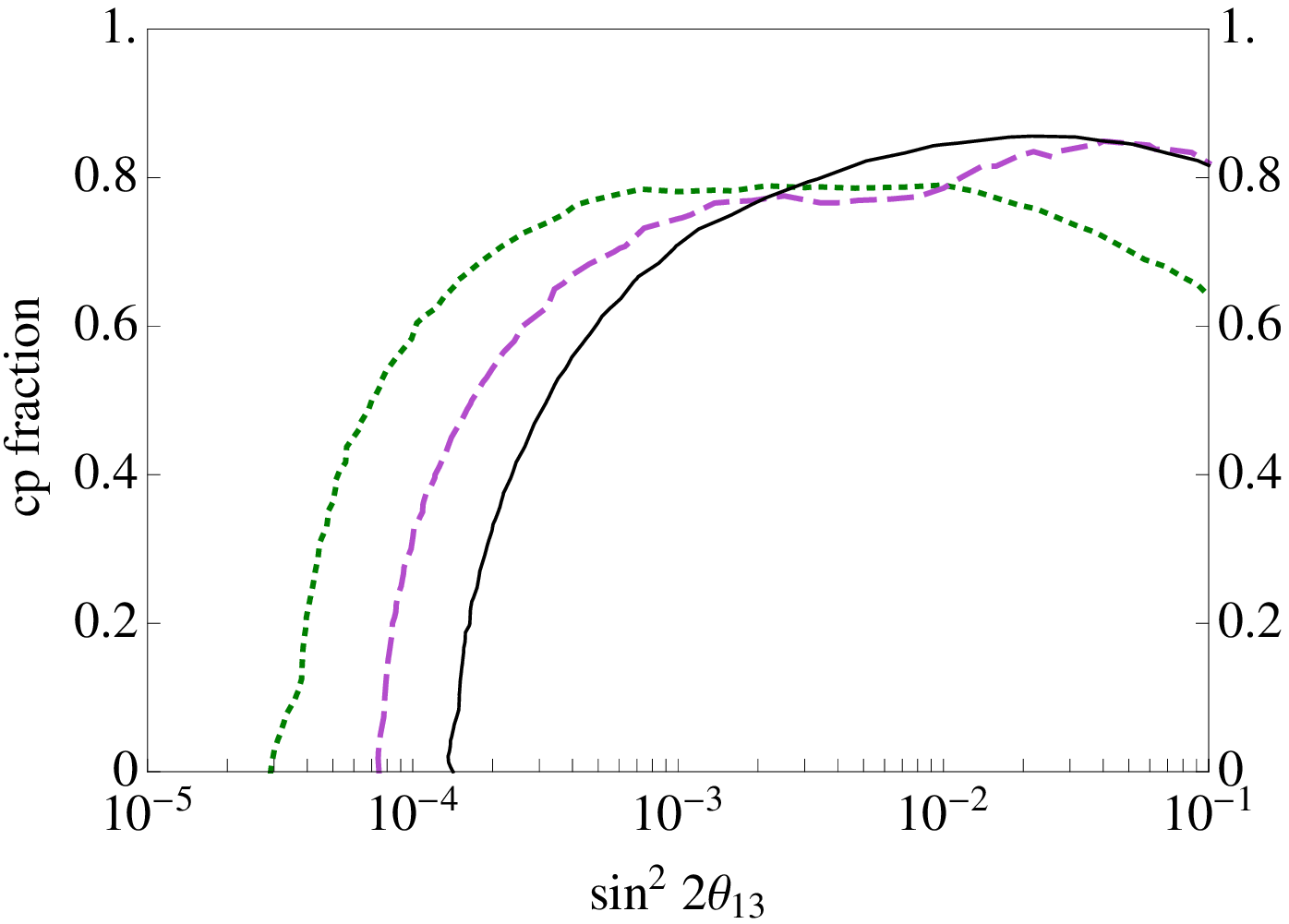}\\
\includegraphics[width=0.40\textwidth,angle=0]{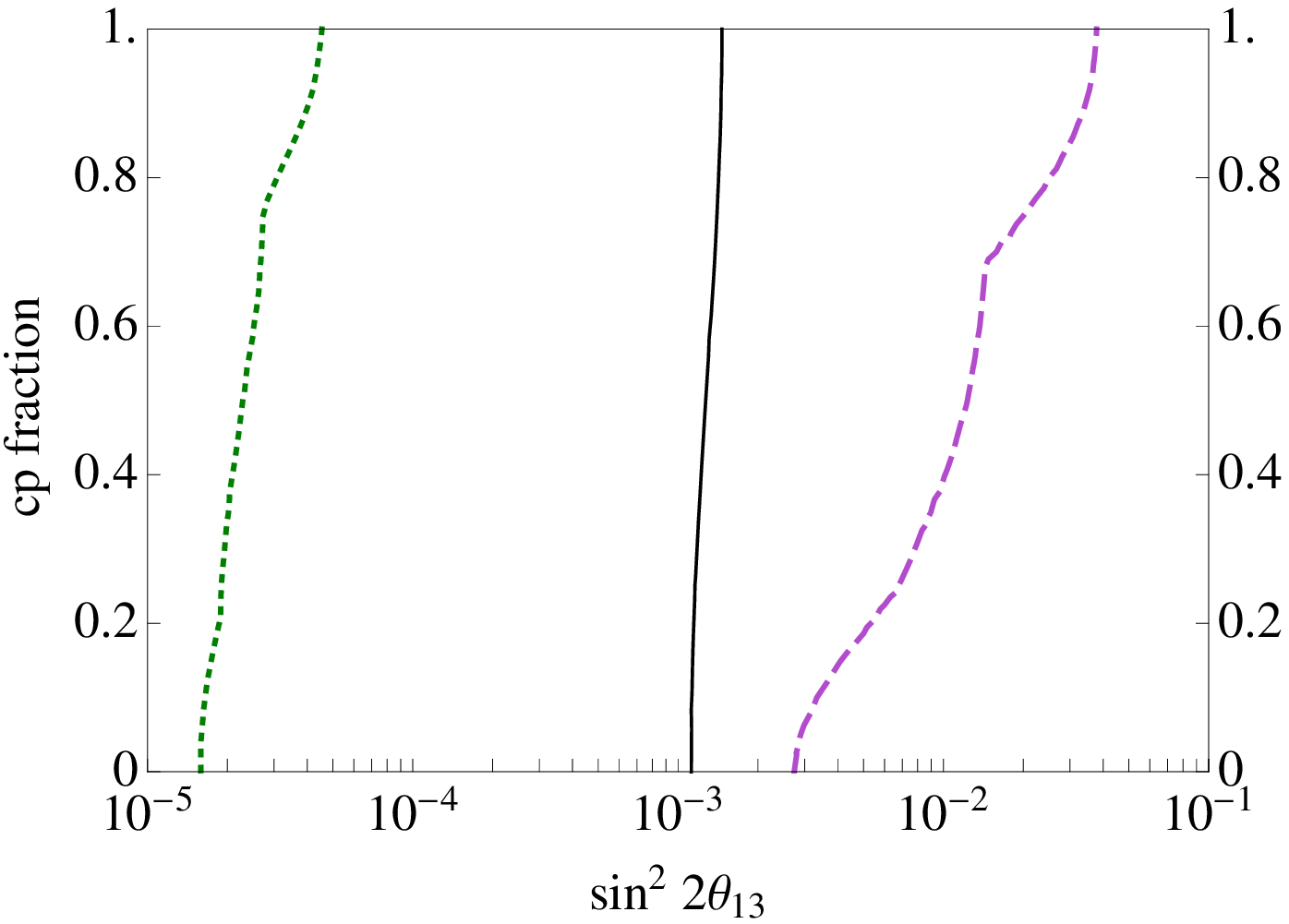}&
\includegraphics[width=0.40\textwidth,angle=0]{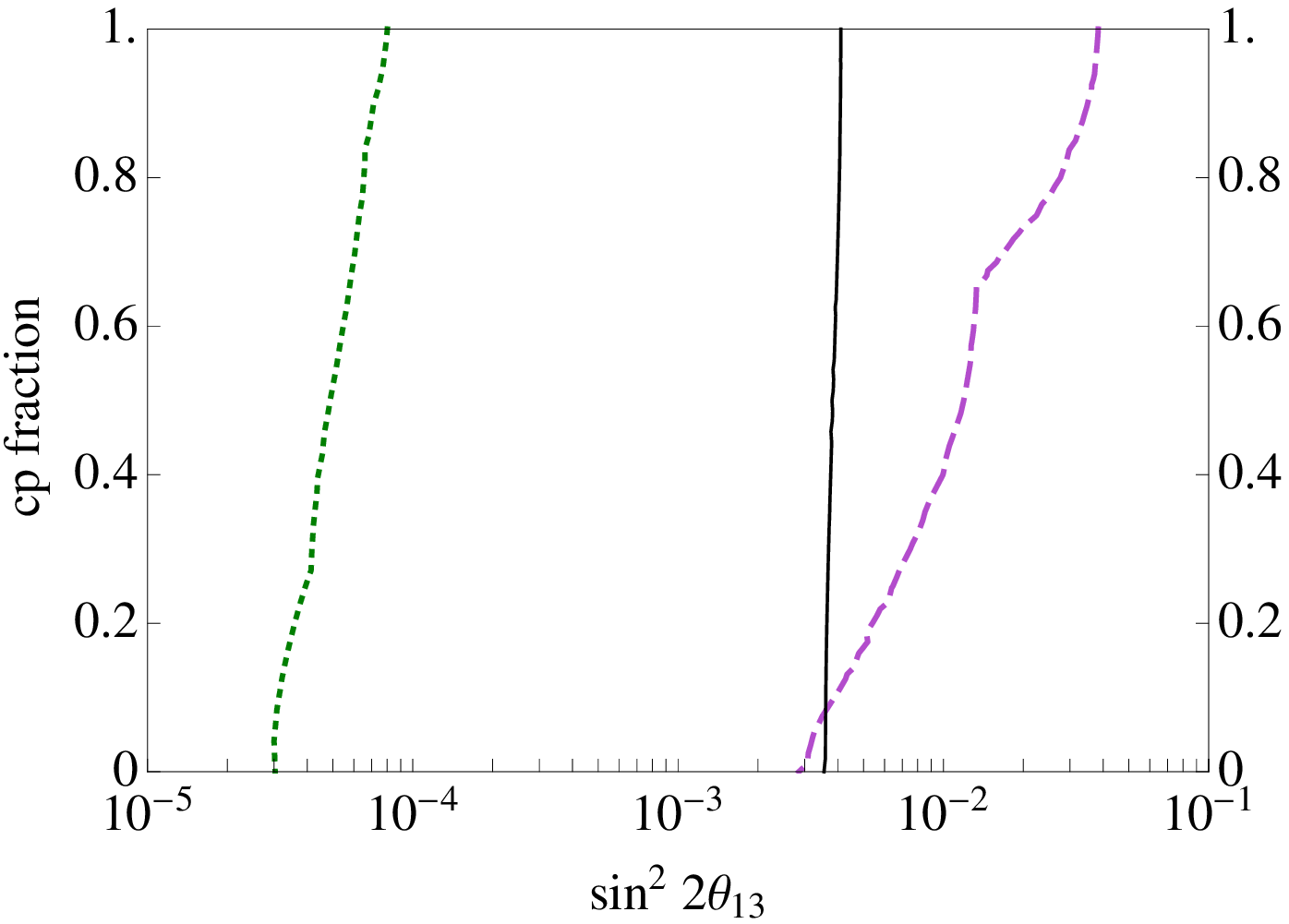}\\
\end{tabular}

\caption{\label{fig:senscpf}
Comparison of our proposed set-up (black solid lines) with the IDS Neutrino Factory baseline design (green dotted lines) and 
the high $\gamma$ \bb set-up from \cite{bc,bc2}. The upper left hand panel shows $\stch$ discovery reach, the upper right hand 
panels shows the CP violation reach, while the lower panels show the mass hierarchy discovery reach for normal (left panel) and inverted (right pannel) hierarchy.  }
\end{figure}

We present in Fig.~\ref{fig:senscpf} the comparison of the performance of our setup with the IDS Neutrino Factory design \cite{Bandyopadhyay:2007kx} and the high-$\gamma$ \bb set-up 4, described above. In this case we present the sensitivities to the observables as a function of the fraction of the values of $\delta$ for which they can be discovered instead of the true values of $\delta$.

From Fig.~\ref{fig:senscpf} it is clear that the facility with sensitivity to the different observables down to smallest values of $\stch$ is the Neutrino Factory. This can be understood from the very large fluxes assumed for the IDS baseline as compared to the ones assumed here for the $\beta$-Beam set-ups: $5\times10^{20}$ useful muon decays per year and per baseline to be compared to the $3 \times 10^{18}$ assumed for the $\beta$-Beams. On the other hand, the high energy of the Neutrino Factory beams implies a very small value of $L/E_\nu$. This translates into a stronger suppression of the CP violating term of the oscillation probability with respect to the one suppressed by two powers of $\theta_{13}$ for large values of this parameter. Therefore, the CP discovery potential of $\beta$-Beams outperforms that of the Neutrino Factory in Fig.~\ref{fig:senscpf} when $\sin^2 2 \theta_{13}~>~10^{-3}$.  Since this large value of $\sin^2 2 \theta_{13}$ also guarantees a discovery of the mass hierarchy and $\stch$ regardless of the value of $\delta$, this makes $\beta$-Beams the better option when $\sin^2 2 \theta_{13}~>~10^{-3}$. Furthermore, even if the statistics in the near $\beta$-Beam detector is reduced by half in our proposal compared to the one in Ref.~\cite{bc,bc2}, the CP-discovery potential for $\sin^2 2 \theta_{13}~>~10^{-3}$ is better in the two-baseline set-up due to the lifting of the degeneracies that can mimic CP-conservation when combining the information from the two detectors.

\section{Conclusions}

We have presented a new $\beta$-Beam set-up that combines the strengths of the best set-ups in the literature, trying to probe with the same facility $\theta_{13}$, the existence of leptonic CP-violation and the neutrino mass ordering in the challenging regime of small $\theta_{13}$.

In order to achieve good sensitivity to the CP phase, we have chosen the highest $\gamma$ accesible at the SPS+ but exploiting the decay of the ions with smallest end-point energy. This guarantees good statistics at the detector, since the flux and cross sections grow with $\gamma$, while maintaining the mean energy around $1$~GeV, which allows to consider a water \chr detector. Furthermore, the oscillation baseline can be kept short, so as to further increase the statistics and to avoid strong matter effects that could spoil the CP discovery potential. 

On the other hand, to achieve sensitivity to the mass hierarchy, a far detector located at the magic baseline is mandatory. For this detector, ions with higher end-point energies are preferred, and higher $\gamma$ values help to increase the statistics. As for this case the neutrino fluxes are peaked around $5-6$~GeV, we opted for an iron detector. 

While two-baseline \bb set-ups have been proposed and studied before, our proposal is unique. We propose two different racetrack
geometry decay rings, as we use different ions for the two baselines. We have kept the initial design for the ring that is aiming at the near detector. However, the \br and \li beam has to be sent over a baseline $L=7000$ km, and hence its storage ring requires an inclination of $\vartheta=34.5^\circ$. This would require a maximum depth $d=2132$ m at the far end of the storage ring if we use the same design. In order to alleviate this problem, one necessarily has to reduce the size of the straight sections of the ring that is aiming to the far detector. To compensate for the consequent loss in the number of events due to this reduction, we propose to increase the $\gamma$ for the \br and \li ions.

We have also made a full comparison with the rest of proposed facilities. While the presently assumed $\beta$-Beam fluxes cannot compete with the expectations from a Neutrino Factory and cannot probe values of $\theta_{13}$ much smaller than $\stch \sim 10^{-4}$, we find that $\beta$-Beam set-ups are better optimized for regions with $\sin^22\theta_{13}~>~10^{-3}$, providing sensitivity to the different observables in larger fractions of the parameter space. We believe that the combination of ions and baselines proposed here represents an optimal \bb set-up, that takes advantage of the properties of the different achievable beams, with very good sensitivity to all of the three observables considered.


\end{document}